\documentclass[journal=jctcce,manuscript=article]{achemso}
\setkeys{acs}{articletitle = true}

\usepackage[version=3]{mhchem} 
\usepackage[T1]{fontenc}       
\usepackage{color,soul}
\usepackage{mhchem}
\usepackage{tikz}
\usepackage{pgfplots}
\usepackage{subcaption}
\usepackage{gensymb}
\usepackage{xr}
\usepackage[compact]{titlesec}
\titlespacing{\section}{0pt}{*0}{*0}

\makeatletter
\newcommand*{\addFileDependency}[1]{
  \typeout{(#1)}
  \@addtofilelist{#1}
  \IfFileExists{#1}{}{\typeout{No file #1.}}
}
\makeatother

\author{Carlos Manuel de Armas-Morej\'on}
\email{carlosdearmasm@gmail.com}

\author{Luis A. Montero-Cabrera}
\email{lmc@fq.uh.cu}
\affiliation[UHavana]
{Laboratorio de Química Computacional y Teórica,
          Facultad de Química, Universidad de La Habana,
          10400. La Habana, Cuba.}
\author{Angel Rubio}
\email{angel.rubio@mpsd.mpg.de}
\affiliation[mpsd]
{Theory Department, Max Planck Institute for the Structure and Dynamics
of Matter and Center for Free-Electron Laser Science, Luruper Chaussee 
149, 22761 Hamburg, Germany}
\affiliation[Nanobio]
{Nano-Bio Spectroscopy Group,  Departamento de Polímeros y Materiales Avanzados: Fisica, Química y Tecnología, Universidad del País Vasco UPV/EHU- 20018 San Sebastián, Spain.}
\author{Joaquim Jornet-Somoza}
\email{j.jornet.somoza@gmail.com}
\affiliation[Nanobio]
{Nano-Bio Spectroscopy Group,  Departamento de Polímeros y Materiales Avanzados: Fisica, Química y Tecnología, Universidad del País Vasco UPV/EHU- 20018 San Sebastián, Spain.}
\alsoaffiliation[mpsd]
{Theory Department, Max Planck Institute for the Structure and Dynamics of 
Matter and Center for Free-Electron Laser Science, Luruper Chaussee 149, 
22761 Hamburg, Germany}

\title[An \textsf{achemso} demo]
  {Electronic Descriptors for Supervised Spectroscopic Predictions}

\abbreviations{ML, KRR, CNN}

\begin{document}

\begin{tocentry}

\includegraphics[width=1.\textwidth]{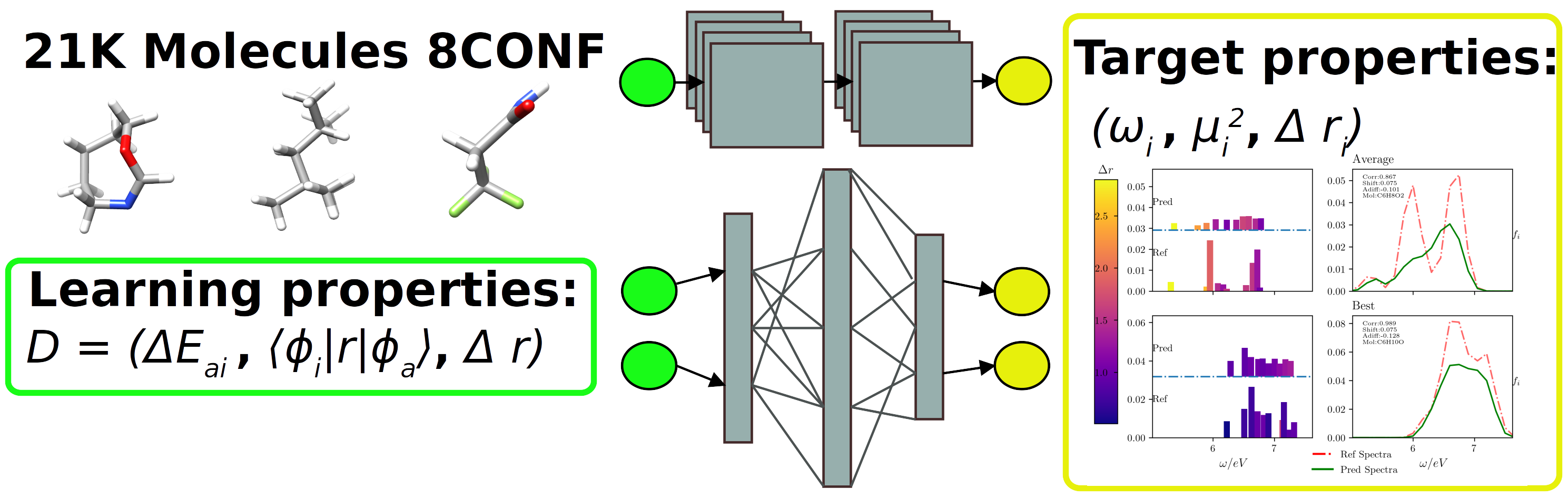}

\end{tocentry}

\begin{abstract}

Spectroscopic properties of molecules holds great importance for the description of the molecular response under the effect of an UV/Vis electromagnetic radiation. Computationally expensive \emph{ab initio} (e.g. MultiConfigurational SCF, Coupled Cluster) or TDDFT methods are commonly used by the quantum chemistry community to compute these properties. In this work, we propose a (supervised) Machine Learning approach to model the absorption spectra of organic molecules. Several supervised ML methods have been tested such as Kernel Ridge Regression (KRR), Multiperceptron Neural Networs (MLP) and Convolutional Neural Networks\cite{ml_6, ml_35}. The use of only geometrical descriptors (e.g. Coulomb Matrix) proved to be insufficient for an accurate training \cite{ml_6}. Inspired on the TDDFT theory, we propose to use a set of electronic descriptors obtained from low-cost DFT methods: orbital energy differences ($\Delta \epsilon_{ia} = \epsilon_a - \epsilon_i$), transition dipole moment between occupied and unoccupied Kohn-Sham orbitals ($\langle \phi_i |r| \phi_a \rangle$) and charge-transfer character of mono-excitations ($R_{ia}$). We demonstrate that with this electronic descriptors and the use of Neural Networks we can predict not only a density of excited states, but also getting very good estimation of the absorption spectrum and charge-transfer character of the electronic excited states, reaching results close to the chemical accuracy ( $\sim 2$ kcal/mol  or $\sim 0.1$ eV).
\end{abstract}

\section{Introduction}

The absorption spectra hold great importance for discovering photo-electric features in material science. The design of new photo-sensitive devices and materials for the energy industry as well as healthcare became a hot topic in the last decades. The increase of the experimental and \emph{ab-initio} theoretical databases\cite{ml_6, ml_125} on materials pushed forward a new way for their discovery and design,  but usually they do not incorporate all the required spectroscopic information. Then, researchers rely on quantum mechanics techniques, usually Time-Dependent Density Functional Theory (TDDFT)\cite{ml_46, ml_108} or multiconfigurational wave-function methods\cite{ml_44}, for an accurate prediction of properties and characterization. However, these type of calculation are usually complex to perform and to understand for non trained researchers, particularly when trying to get reliable predictions of absorption spectra from an initial selection within several candidates.

Recently, Machine Learning (\textbf{ML}) algorithms have attracted the interest of the research community because the plauseible results obtained predicting materials properties with good accuracy.\cite{ml_6, ml_21}  ML algorithms have been used, for example, for property classifications and group discovery\cite{ml_5, ml_3, ml_47}, as well as ground-state material and molecular property predictions\cite{ml_1, ml_21, ml_49, ml_70}.

A profound research on several ML methods to be chosen, such as supervised or unsupervised models, kernel regression methods or neural networks, etc, is required for each type of target property. Moreover, the choice of the appropriate molecular descriptors has to be made carefully in order to fulfill some desired criteria: 1) simplicity: must be easy to produce, 2) representability: must contain the required information correlated to the target property and 3) specificity: must be unique enough to distinguish between different molecules.\cite{ml_4} Several descriptors have been proposed in the literature with different levels of applications. \cite{ml_3, ml_11, ml_12, ml_13, ml_10, ml_34, ml_56, ml_79, ml_92, ml_93, ml_94, ml_95, ml_9, ml_97, ml_98, ml_99, ml_130, ml_131, ml_132, ml_133, ml_134, ml_135, ml_136, ml_136, ml_137, ml_138, ml_139, ml_140, ml_141, ml_142, ml_143, ml_144, ml_145} Ouyang et al. \cite{ml_3}  propose also the SISSO method for construct these molecular or material descriptors base on algebraic combinations of atomic properties.

Several attempts have been made to predict theoretical spectroscopic properties for molecules\cite{ml_6, ml_70} and materials. \cite{ml_2, ml_49, ml_11, ml_4} The seminal work done by Ramakrishnan et al.\cite{ml_6} proposed a kernel ridge regression model that can predict the first excited state with good results. Besides they proposed a method called $\Delta ML$ for the estimation of the shift between two databases obtained using different Exchange-Correlation (XC) functionals. In that work, the authors used the so called Coulomb Matrix\cite{ml_21} as a geometrical descriptor, which has gained notoriety because of its low computational requirements and its good performance for predicting molecular properties.\cite{ml_6, ml_35} However, it proved to be insufficient for the proper prediction of the transition probability. \cite{ml_6} In a recent work, Westermayr et al. found a machine learning model based on the use of a complex Neural Network  that using many conformers of the same molecule as a training set can be used to accurately predict its absorption spectra.\cite{ml_68}  

In this work, we propose for the first time the use of some calculated electronic properties in order to well characterize the spectroscopic fingerprint of small molecules. By using a simple Convolutional Neural Network model trained by low-cost theoretical electronic calculations obtained from a 21k molecular database we can predict excitation energies with the corresponding charge transfer character and oscillator strength of small molecules. The results presented in this paper are obtained by employing electronic descriptors from ground-state DFT calculations using a LDA XC-functional to predict the absorption spectra at a TDDFT level using the PBE0 hybrid XC-functional. The validity of the selected model is contrasted with different Neural Network schemes, and the limitations are described on the base of obtained results. 

\subsection{Molecular database and Descriptor Selection.}\label{sec:mol_desc}

In this work, we use a subset of the \emph{GDB-8} molecular database also used by Ramakrishnan et al. \cite{ml_6, ml_125, ml_126}. It consists of 21k small organic molecules with relaxed geometries as computed by using DFT at the B3LYP|6-31(2df,p) level. The selected molecules contain up to 8 carbon (C), oxygen (O), nitrogen (N) and/or fluor (F) atoms, being the number of Hydrogen atoms the required to make neutral the molecular charges. Hereon we will refer it as the 8CONF database.

Although the Sorted Coulomb Matrix and its variants have previously shown good results for the prediction of excitation energy levels and density of states \cite{ml_6}, the use of only geometrical molecular descriptors proved to be insufficient for the correct prediction of transition moments or oscillator strengths. 

For that reason, we propose to use an electronic molecular descriptor from low-cost theoretical calculations (ground-state LDA) to predict accurate spectroscopic properties computed at TDDFT level with an hybrid exchange-correlation functional (TDDFT-PBE0).

The choice of the electronic descriptor has been made regarding the linear-response time-dependent DFT formulation (\textbf{LR-TDDFT}). \cite{ml_46, ml_108} This approach aims to solve the time-dependent Shr\"{o}dinger equation,

\begin{equation}\label{eq:time_dep_shro}
 \hat{H}(t)\varPsi(t) = i\frac{\partial \varPsi(t)}{\partial t}, \quad
 \hat{H}(t) = \hat{T} + \hat{V}_{ee} + \hat{V}_{ext}(t)
\end{equation}

where $\hat{H}$ is the system Hamiltonian composed by a kinetic part ($\hat{T}$), an electron-electron potential component ($\hat{V}_{ee}$), and the all-other type of interaction contained in the time-dependent external potential term $\hat{V}_{ext}(t)$. Usually, the latter contains the nuclei-electron and external field interaction.

In LR-TDDFT the time-dependent evolution of the non-interacting system under an external field is described by the non-interacting \emph{density-density} response function $\chi_{s} (\boldsymbol{r},\boldsymbol{r}',\omega)$,  

\begin{equation}\label{eq:response_chi}
\chi_{s} (\boldsymbol{r},\boldsymbol{r}',\omega) = \lim_{\eta\to0^+} \sum_{k,j}(f_k-f_j)\delta_{\sigma_k\sigma_j} \frac{\varphi_k^{(0)*}(\boldsymbol r)\varphi_j^{(0)}(\boldsymbol r)\varphi_j^{(0)*}(\boldsymbol r')\varphi_k^{(0)}(\boldsymbol r')}{\omega - (\epsilon_j-\epsilon_k) + i\eta}
\end{equation}

where $\varphi_i$ stands for a Kohn-Sham orbital, $\epsilon_j$ and $f_i$ the respective corresponding energy and occupation, and $\delta_{\sigma_k\sigma_j}$ is a Kronecker delta orbital $j$ and $k$ spin functions.$\omega$ is the frequency of the perturbing external field and $\eta$ is a positive infinitesimal.

This function has poles on the excitation energy of the KS system. In order to obtain the excitation energies of the full interacting system we have to solve the Dyson like equation. Casida et al. proposed a matrix formulation to solve this equation and he obtained the well-known Casida's equation \cite{ml_48}. By solving this equation, the excitation energy levels and oscillator strengths are obtained as a combination of the bi-orbital function $\Phi_{ia}(\boldsymbol r) = \varphi_i^*(\boldsymbol r)\varphi_a(\boldsymbol r)$ where subindices $a$ and $i$ correspond to unoccupied and occupied states respectively.

In the present work, we use the excitation energies and oscillator strengths of $15$k molecules computed using the Casida's equations at the PBE0 functional level in the ground-state, both to train our ML model and as targets to validate it. 

Let us return to the LR-TDDFT formulation in order to define the descriptors we will use. The time-evolution of the polarizability function is defined n LR-TDDFT as the dipole-dipole response function, which in the space of frequencies takes the form, 

\begin{equation} \label{eq:alpha}
    \alpha_{\mu\lambda}(\omega) = \sum_{n=1}^{\infty}\Bigg \{ \frac{ 2\Omega_n \langle \Psi_0 |\hat r_\nu| \Psi_n\rangle \langle \Psi_n | \hat r_\lambda| \Psi_0\rangle}{\Omega_n^2 - \omega^2}  
\Bigg \} 
\end{equation}

where, $\Psi_0$ is the ground-state wave-function, $\Psi_n$ is the $n$-th excited state wave-function, $\Omega_n$ is the excitation energy of the $n$-th excited state, $\mu,\lambda$ are directions in the space, and $\hat r_{\nu}$ is the dipole operator in the $\nu$ direction. 

This function has also poles in the excitation energies of the system and the corresponding oscillation strengths are proportional to the numerator. Therefore, it is then also used to compute the absorption spectra of molecular systems. 

Based on equation \ref{eq:alpha} and the use of bi-orbital functions representing monoelectronic excitations, we propose to use the following electronic descriptors:

\begin{enumerate}
    \item  Orbital energy difference: $\Delta \epsilon_{ia} = \epsilon_{a} - \epsilon_{i}$.
    \item  Kohn-Sham transition moment: $\mu_{ia}^2 = | \langle \varphi_i | \boldsymbol{r} | \varphi_a \rangle |^2$
\end{enumerate}

Nevertheless, the only use of these two properties does not fulfill the desired criteria of specificity described above. The calculated oscillator strengths depend on orbital overlapping between unoccupied and occupied states, and are hence proportional to transition dipole moments. Besides, a work of Guido et al.  \cite{ml_34} proposes an easy way to evaluate the charge transfer character of an excitation by defining a new index, $\Delta \boldsymbol{r}$:

\begin{equation}\label{eq:char_tranf}
\varDelta \boldsymbol{r} =
    \frac{\sum\limits_{i, a}{K_{ia}^2 R_{ia}}}{\sum\limits_{i,a}{K_{ia}^2}}
\end{equation}
\begin{equation}
    R_{ia} = |\langle \varphi_a | \boldsymbol{r} | \varphi_a \rangle - \langle \varphi_i  | \boldsymbol{r} | \varphi_i \rangle| .
\end{equation}
\begin{equation}\label{eq:td-coeff}
K_{ia} = X_{ia} + Y_{ia}
\end{equation}

where intervene the excitation $X_{ia}$ and de-excitation $Y_{ia}$ coefficients of the non-Hermitean solution corresponding to the TD formalism.

\begin{figure}[ht!]
{\includegraphics[scale=0.7]
        {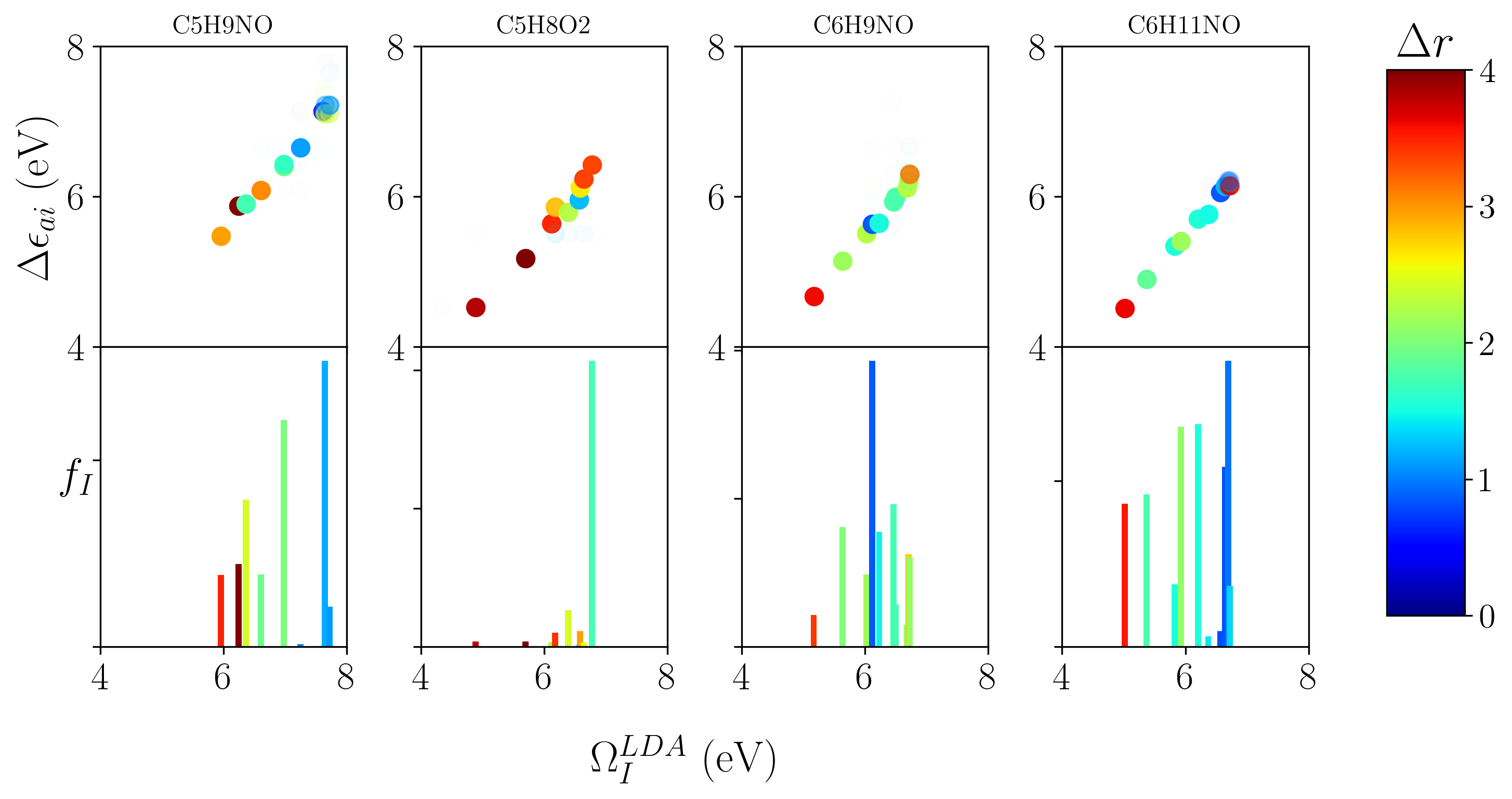}}
\caption{Example for four selected molecules from the 8CONF of correlation between orbital energy differences and LR-TDDFT calculated excitation energies $\Omega^{LDA}_I$ (First row from top to down). The second row shows the calculated discrete absorption spectra for those molecules corresponding to such excitation energies. Transparency is proportional to the Casida's coefficient, and color to the charge transfer character.}
\label{fig:corr_examp}
\end{figure}

Following that index and knowing that only symmetrically similar mono-excitation contributes to the real excited state (Figure \ref{fig:corr_examp}), we decided to also include the charge transfer character of the KS-monoexcitation as a descriptor, as well as the TDDFT charge transfer index of the excited state as a target property to predict. 

Consequently, in this work we propose the use of the combination of three electronic descriptors, namely i) $\Delta \epsilon_{ia}$, ii) $\mu^2_{ia}$ and iii) $R_{ia}$, computed at ground-state LDA XC-functional level and LCAO (LDA), for the twenty lowest-lying mono-electronic transitions to predict three spectroscopic properties for the first ten excitation: a) excitation energy ($\Omega_I$), b) oscillator strength ($fosc_I$) and c) charge transfer character ($\Delta{\boldsymbol{r}}$) at the PBE0 accuracy level.

\subsection{Supervised Machine Learning models}

In this work we use Neural Networks (NN) because their recognised versatility to find hidden correlations between properties. We explore different NN models such as the Multi-Layer Perceptron (MLP) and the Convolution Neural Network (CNN). Each model depends on a group of internal variables known as hyper-parameters such as the number of hidden layers, the number of neurons per layer, the number of learning iterations (also known as \emph{epochs}), the activation functions and many other. Some of these variables need to be optimised to fine-tune the NN.
Hyper-parameter optimizations hold great importance in the correct behaviour of the NN. The values can be empirically selected, but the best combination can be only achieved by a systematic search. Then, we applied a Bayesian Optimization as implemented in \emph{scikit-learn}\cite{ml_65} to find the optimal values for the \emph{number of hidden layers} and the \emph{number of epoch}.

Another important issue is how to feed the data to the NN. The flexibility of the NN allows many configurations for introducing the descriptors into the model. Consider  $d_{m,n}$ an element of the input tensor $\boldsymbol{D}$, where $n$ is the descriptor property ($\Delta \epsilon_{ia}$, $\mu^2_{ia}$ or $R_{ia}$)  and $m$ the considered  $a\leftarrow i$ mono-excitatiton label. We used two strategies to introduce our data into the neural network: (1) The \textbf{1D model}, where each sample $j$ is described by an array of dimension 1 where all properties are introduced sequentially: $D_{j} = ( \Delta \epsilon_1, \mu^2_1, R_1, ..., \Delta \epsilon_m, \mu^2_m , R_m )$; (2)  The \textbf{2D model}, where descriptor properties for each $m$ mono-excitation is grouped in then array, since all these properties are required to describe a particular excitation: $D_{j} = ( (\Delta \epsilon_1, \mu^2_1, R_1), ..., (\Delta \epsilon_m, \mu^2_m , R_m ))$.

The use of different properties, units and ranges of magnitude may affect the learning process. 
It is always recommended to perform a data pre-processing in order to give the same weight for all properties and hence to ease the learning process. In this work, we decided to scale all the data between $[0, 1]$ using the tool \emph{MinMaxScaler} provided by the \emph{preprocessing} package of \emph{scikit-learn}.\cite{ml_65} 
Since the  range of transition dipole moments is always positive and presents a high density distribution for values between 0 and 1, we transformed this property to a logarithmic scale. 

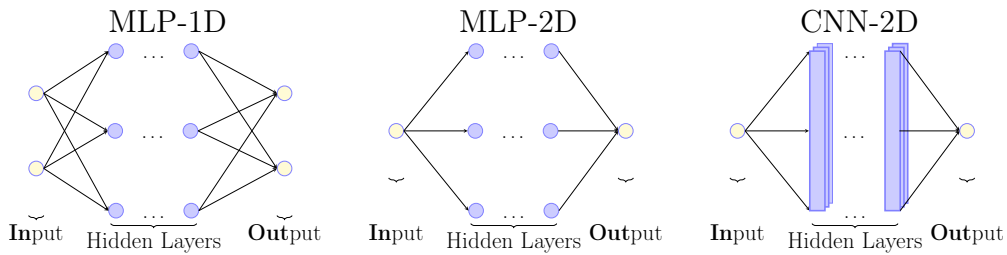
\begin{figure}[ht!]
\centering
\begin{tabular}[c]{ c  c  c }
MLP-1D & MLP-2D & CNN-2D \\
{\scalebox{0.5}{%
    \begin{tikzpicture}[,node distance=3cm]
\node (a) at (0,0) [circle, draw=blue!50,  fill=yellow!20, ]{};
\node (guide)[below of=a, yshift=2cm,
                             ]{};
\node (a1) [below of=a, yshift=1.cm,  circle, draw=blue!50, fill=yellow!20,]{};

\node (b1) [above right of=guide, circle, draw=blue!50, fill=blue!20,]{};
\node (b2) [right of=guide, circle, draw=blue!50, fill=blue!20, xshift=-.9cm,]{};
\node (b3) [below right of=guide, circle, draw=blue!50, fill=blue!20,]{};

\node (b4) [ right of=b1,  circle, draw=blue!50, fill=blue!20, xshift=-1cm]{};
\node (b5) [right of=b2,  circle, draw=blue!50, fill=blue!20, xshift=-1cm]{};
\node (b6) [right of=b3,  circle, draw=blue!50, fill=blue!20, xshift=-1cm]{};

\node (d)  [right of=b5, yshift=1cm, circle, draw=blue!50, fill=yellow!20, xshift=-.5cm, ]{};
\node (d1)  [right of=b5, yshift=-1cm, circle, draw=blue!50, fill=yellow!20, xshift=-.5cm,]{};


\draw [->, >=stealth](a.east)  -- node[anchor=east] {}  (b1.west);
\draw [->, >=stealth](a.east)  -- node[anchor=west] {}  (b2.west);
\draw [->, >=stealth](a.east) -- node[anchor=east] {} (b3.west);

\draw [->, >=stealth](a1.east)  -- node[anchor=east] {} (b1.west);
\draw [->, >=stealth](a1.east)  -- node[anchor=west] {} (b2.west);
\draw [->, >=stealth](a1.east) -- node[anchor=east, yshift=-0.3cm] {} (b3.west);
\draw [draw=none](b1.east) -- node[anchor=north] {\large $\hdots$} (b4.west);
\draw [draw=none](b2.east) -- node[anchor=north] {\large $\hdots$} (b5.west);
\draw [draw=none](b3.east) -- node[anchor=north] {\large $\hdots$} (b6.west);
\draw [->, >=stealth](b4.east)  -- node[anchor=west] {}  (d.west);
\draw [->, >=stealth](b4.east)  -- node[anchor=west] {}  (d1.west);
\draw [->, >=stealth](b5.east) -- node[anchor=west] {} (d.west);

\draw [->, >=stealth](b5.east)  -- node[anchor=west] {} (d1.west);
\draw [->, >=stealth](b6.east)  -- node[anchor=west] {} (d.west);
\draw [->, >=stealth](b6.east) -- node[anchor=west] {} (d1.west);

\draw [decorate, decoration={brace, mirror,raise=0.2cm, amplitude=3pt}]
       (b3.south west)--node[anchor=north, label=below:\Large  Hidden Layers]{}(b6.south east);
\draw [decorate, decoration={brace, mirror, raise=1.25cm,  amplitude=3pt}]
       (a1.west)--node[anchor=north, yshift=-1.05cm,
       label=below:\Large  \textbf{In}put]{}(a1.east);
\draw [decorate, decoration={brace, mirror, raise=1.25cm,  amplitude=3pt}]
       (d1.west)--node[anchor=north, yshift=-1.05cm,
       label=below:\Large \textbf{Out}put]{}(d1.east);
\end{tikzpicture}}} &
{\scalebox{0.5}{%
    \begin{tikzpicture}[,node distance=3cm]
\node (a) at (0,0) [circle, draw=blue!50,  fill=yellow!20 ]{};
\node (guide)[below of=a, yshift=3cm,
                             ]{};
\node (b1) [above right of=guide, circle, draw=blue!50, fill=blue!20,]{};
\node (b2) [right of=guide, circle, draw=blue!50, fill=blue!20, xshift=-.9cm,]{};
\node (b3) [below right of=guide, circle, draw=blue!50, fill=blue!20,]{};

\node (b4) [ right of=b1,  circle, draw=blue!50, fill=blue!20, xshift=-1cm]{};
\node (b5) [right of=b2,  circle, draw=blue!50, fill=blue!20, xshift=-1cm]{};
\node (b6) [right of=b3,  circle, draw=blue!50, fill=blue!20, xshift=-1cm]{};

\node (d)  [right of=b5, yshift=0cm, circle, draw=blue!50, fill=yellow!20, xshift=-1cm, ]{};
\draw [->, >=stealth](a.east)  -- node[anchor=east] {}  (b1.west);
\draw [->, >=stealth](a.east)  -- node[anchor=west] {}  (b2.west);
\draw [->, >=stealth](a.east) -- node[anchor=east] {} (b3.west);

\draw [draw=none](b1.east) -- node[anchor=north] {\large $\hdots$} (b4.west);
\draw [draw=none](b2.east) -- node[anchor=north] {\large $\hdots$} (b5.west);
\draw [draw=none](b3.east) -- node[anchor=north] {\large $\hdots$} (b6.west);
\draw [->, >=stealth](b4.east)  -- node[anchor=west] {}  (d.west);
\draw [->, >=stealth](b5.east) -- node[anchor=west] {} (d.west);

\draw [->, >=stealth](b6.east)  -- node[anchor=west] {} (d.west);

\draw [decorate, decoration={brace, mirror,raise=0.2cm, amplitude=3pt}]
       (b3.south west)--node[anchor=north, label=below:\Large Hidden Layers]{}(b6.south east);
\draw [decorate, decoration={brace, mirror, raise=1.25cm,  amplitude=3pt}]
       (a.west)--node[anchor=north, yshift=-2.05cm,
       label=below:\Large \textbf{In}put]{}(a.east);
\draw [decorate, decoration={brace, mirror, raise=1.25cm,  amplitude=3pt}]
       (d.west)--node[anchor=north, yshift=-2.05cm,
       label=below:\Large \textbf{Out}put]{}(d.east);
\end{tikzpicture}}} &
{\scalebox{0.5}{%
    \begin{tikzpicture}[,node distance=3cm]
\node (a) at (0,0) [circle, draw=blue!50,  fill=yellow!20 ]{};
\node (guide)[below of=a, yshift=3cm,
                             ]{};
\node (b1) [above right of=guide, circle, draw=white, fill=white,]{};
\node (b2) [right of=guide, circle, draw=white, fill=white, xshift=-.9cm,]{};
\node (b3) [below right of=guide, circle, draw=white, fill=white,]{};

\node (b4) [ right of=b1,  circle, draw=white, fill=white, xshift=-1cm]{};
\node (b5) [right of=b2,  circle, draw=white, fill=white, xshift=-1cm]{};
\node (b6) [right of=b3,  circle, draw=white, fill=white, xshift=-1cm]{};

\node (d)  [right of=b5, yshift=0cm, circle, draw=blue!50, fill=yellow!20, xshift=-1cm, ]{};
\draw [->, >=stealth](a.east)  -- node[anchor=east] {}  (b1.west);
\draw [->, >=stealth](a.east)  -- node[anchor=west] {}  (b2.west);
\draw [->, >=stealth](a.east) -- node[anchor=east] {} (b3.west);

\draw [draw=blue!50, fill=blue!20, transform canvas={xshift = 0.2cm, yshift=0.2cm, very thick}] (b1.west) rectangle (b3.east);
\draw [draw=blue!50, fill=blue!20, transform canvas={xshift = 0.1cm, yshift=0.1cm}, very thick] (b1.west) rectangle (b3.east);
\draw [draw=blue!50, fill=blue!20, very thick] (b1.west) rectangle  (b3.east);

\draw [draw=blue!50, fill=blue!20, transform canvas={xshift = 0.2cm, yshift=0.2cm}, very thick] (b4.west) rectangle (b6.east);
\draw [draw=blue!50, fill=blue!20, transform canvas={xshift = 0.1cm, yshift=0.1cm}, very thick] (b4.west) rectangle (b6.east);
\draw [draw=blue!50, fill=blue!20, very thick] (b4.west) rectangle (b6.east);

\draw [draw=none](b1.east) -- node[anchor=north] {\large $\hdots$} (b4.west);
\draw [draw=none](b2.east) -- node[anchor=north] {\large $\hdots$} (b5.west);
\draw [draw=none](b3.east) -- node[anchor=north] {\large $\hdots$} (b6.west);
\draw [->, >=stealth](b4.east)  -- node[anchor=west] {}  (d.west);
\draw [->, >=stealth](b5.east) -- node[anchor=west] {} (d.west);

\draw [->, >=stealth](b6.east)  -- node[anchor=west] {} (d.west);

\draw [decorate, decoration={brace, mirror,raise=0.2cm, amplitude=3pt}]
       (b3.south west)--node[anchor=north, label=below:\Large Hidden Layers]{}(b6.south east);
\draw [decorate, decoration={brace, mirror, raise=1.25cm,  amplitude=3pt}]
       (a.west)--node[anchor=north, yshift=-2.05cm,
       label=below:\Large \textbf{In}put ]{}(a.east);
\draw [decorate, decoration={brace, mirror, raise=1.25cm,  amplitude=3pt}]
       (d.west)--node[anchor=north, yshift=-2.05cm,
       label=below:\Large \textbf{Out}put]{}(d.east);
\end{tikzpicture}}} \\
\end{tabular}
\caption{Simplified view of the topology builds on top of \emph{Keras}\cite{keras2015} and \emph{Tensorflow}\cite{tf2015}. From left to right MLP-1D, where the input data is treated independently, MLP-2D and CNN-2D where the input data is treated as a combination of descriptors in both cases. See text.}
\label{fig:nn_models}
\end{figure}

Alongside with the pre-processing methods several NN models have been tested to find which will best perform for predicting properties. Figure \ref{fig:nn_models} represents the different ML models tested in this work. 

We construct our models using Keras and TensorFlow \cite{tf2015, keras2015} with the hyper-parameters shown in Table \ref{tab:nn_config} resulting from a Bayesian Optimizations against the accuracy values obtained over the test set.  Since the number of hyper-parameters is large, only two of them have been optimized: 1) the number layers, and 2) the number of epochs. The activation function, which transforms the values between neurons, has been selected empirically resulting from the use of the eLU\cite{ml_38} and ReLU\cite{ml_39} functions. Both activation functions were alternated starting with eLU. The full description of the models used in this work can be found in the Supplementary Information.

\begin{table}[!ht]
\caption{Relevant hyper-parameters used to build our models. The second and third columns show the number of \emph{Epochs} and \emph{ Hidden Layers}, respectively, as optimised for the model shown in the first column. The fourth column specifies the type of activation function and properties.  For a complete list see Supplementary Information. 
}\label{tab:nn_config}
\begin{center}
\begin{tabular}{ c c c c}
    \hline
    Model &  Epochs & N. Hidden Layers & Activation Function\\
    \hline
    MLP-1D  & 1756 & 17 & eLU/ReLU(negative slope = 0.01) \cite{ml_38, ml_39}\\
    MLP-2D   & 1419 &  4 & eLU/ReLU(negative slope = 0.01) \cite{ml_38, ml_39}\\
    CNN-2D  & 1500 &  2 & eLU/ReLU\\
\end{tabular}
\end{center}
\end{table}

By using this configuration, the training process with 10k molecules as a learning set and the prediction of 1k molecules from the test set takes $\sim 1528$ s for the slowest NN (CNN-2D model) on a commercial laptop with an Intel{\textregistered} i7 and $12$ Gb of RAM.

Table \ref{tab:nn_times} shows the required times for the training and predicting processes. It is important here to remark that once the model is trained, the prediction of the spectroscopic properties are almost instantaneously obtained. Comparing this performance with the arduous task of solving the complex TDDFT equations shows the potential of using neural networks. 

\begin{table}[!ht]
\caption{Approximately \emph{Training} and \emph{Predicting} times in seconds (s) for each model using $10$k molecules and $1$k molecules respectively. Even if CNN-2D takes $~1521$ s to train, which is the slowest one, the response when predicting the properties is almost instantaneous.} \label{tab:nn_times}
\begin{center}
\begin{tabular}[c]{ccccc}
    \hline
    Model Name &  Use $log$ & Training Process (s) & Predicting Process (s) &
    Total Time (s)\\
    \hline
    MLP-1D & Yes &  919  & <1 & 919  \\
    MLP-2D  & Yes &  319  & <1 & 319  \\
    CNN-2D  & Yes & 1521  & <1 & 1522 \\
\end{tabular}
\end{center}
\end{table}

The learning process is obviously biased by the learning data set. In order to validate the  input data distribution, the optimized model and its uncertainty, Musi et al. proposed to systematically perform a stability test.\cite{ml_33}
The learning data set is validated by repeating the neural network construction process (training and predicting process) over $10$ experiments,  by randomly selecting $10k$ learning molecules out of $~15$k PBE0 available calculations. We used $1k$ of the remaining $3k$ molecules as control, which will remain unchanged across all experiments.

\begin{figure}[!ht]
    \centering
    {\includegraphics[scale=0.5]{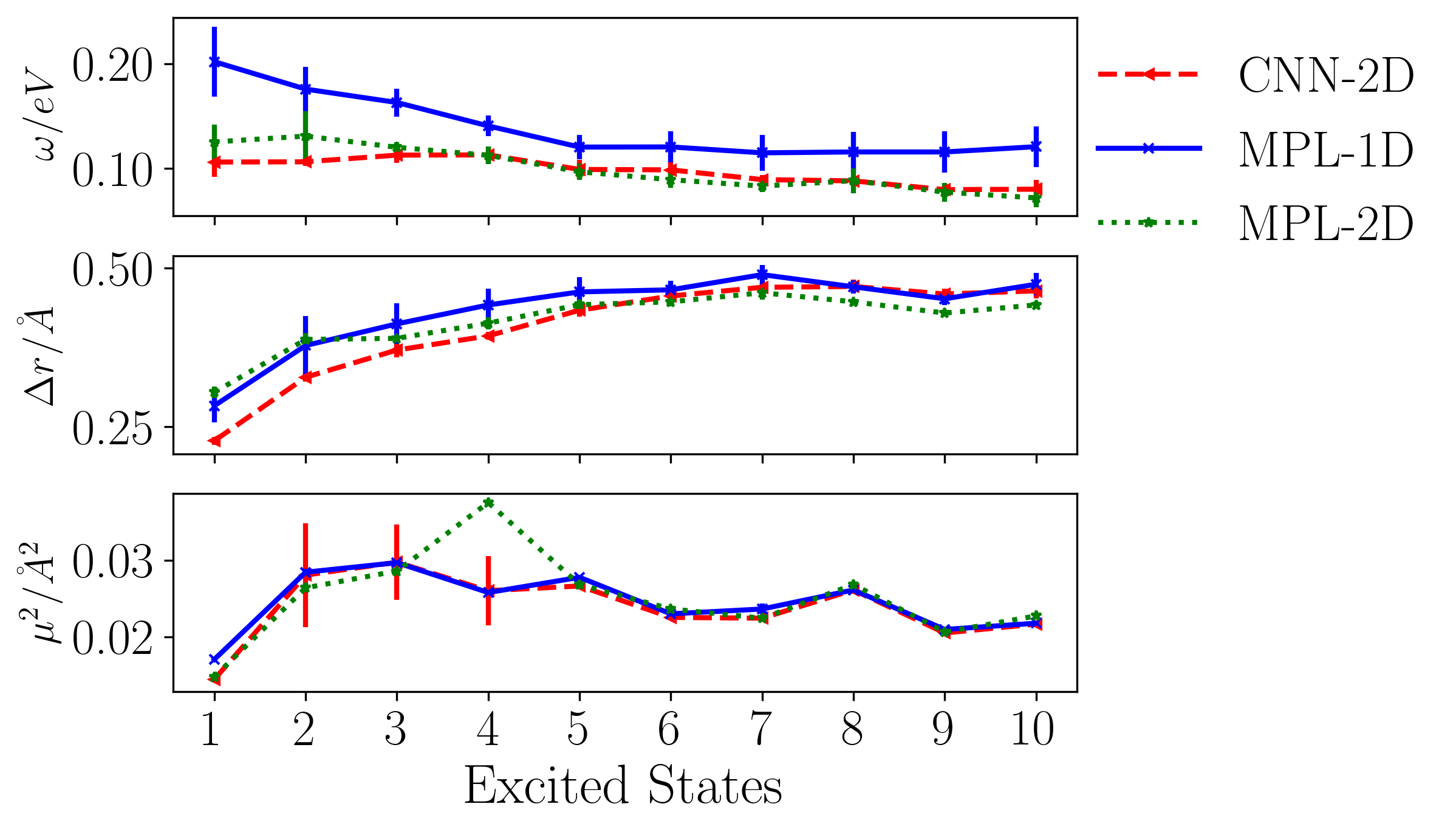}}
    \caption{Mean absolute error (MAE) on the prediction of the excited state energies $\Omega$ (top), charge transfer character index $\Delta{\boldsymbol{r}}$ (middle)  and  transition moments $\mu^2$ (down) for 10 low-laying states averaged over the 10 repetitions of the same ML model. The MAE standard deviation of those experiments is also represented by the bars amplitude.}
    \label{fig:models_stress_curves}
\end{figure}

Figure \ref{fig:models_stress_curves} shows the mean average errors (MAE) for the first 10 excited states of 1k 8CONF small molecules used in the control set. For each state the mean value of the error over the 10 experiments is represented and its standard deviation is depicted as a bar line. 

We can see that  CNN-2D and MPL-1D models reach errors of the excitation energy predictions that are close to the chemical accuracy ($\sim 0.1$ eV). Besides, they can also correctly predict the charge-transfer character of the low-lying excited state. Notice that Figure 1 shows $\Delta \boldsymbol{r}$ property as ranging from $0$ to $4$ \r{A}. It means that a resulting error smaller being than $0.50$ \r{A} clearly distinguishes between short and long range charge transfer characters. Regarding the transition moment prediction, we see that the models can just fairly predict the first excitation probability. 

In the following section we discuss the spectroscopic properties obtained using the more promising models and we will remark their strength and drawbacks.

\section{Results and discussion}

As already described above, the main goal of this work is to find an adequate molecular-electronic descriptor that enables us to obtain accurate spectroscopic properties using machine learning techniques. From the analysis of accuracy and stability (Figure \ref{fig:models_stress_curves}),  we see that CNN-2D and MLP-2D produce the lowest error for predictions. Therefore, we selected these models to perform a deeper analysis.

\begin{figure}[!ht]
\begin{tabular} {c c}
(a) CNN-2D & (b) MLP-2D  \\ 
{\includegraphics[scale=0.5]
{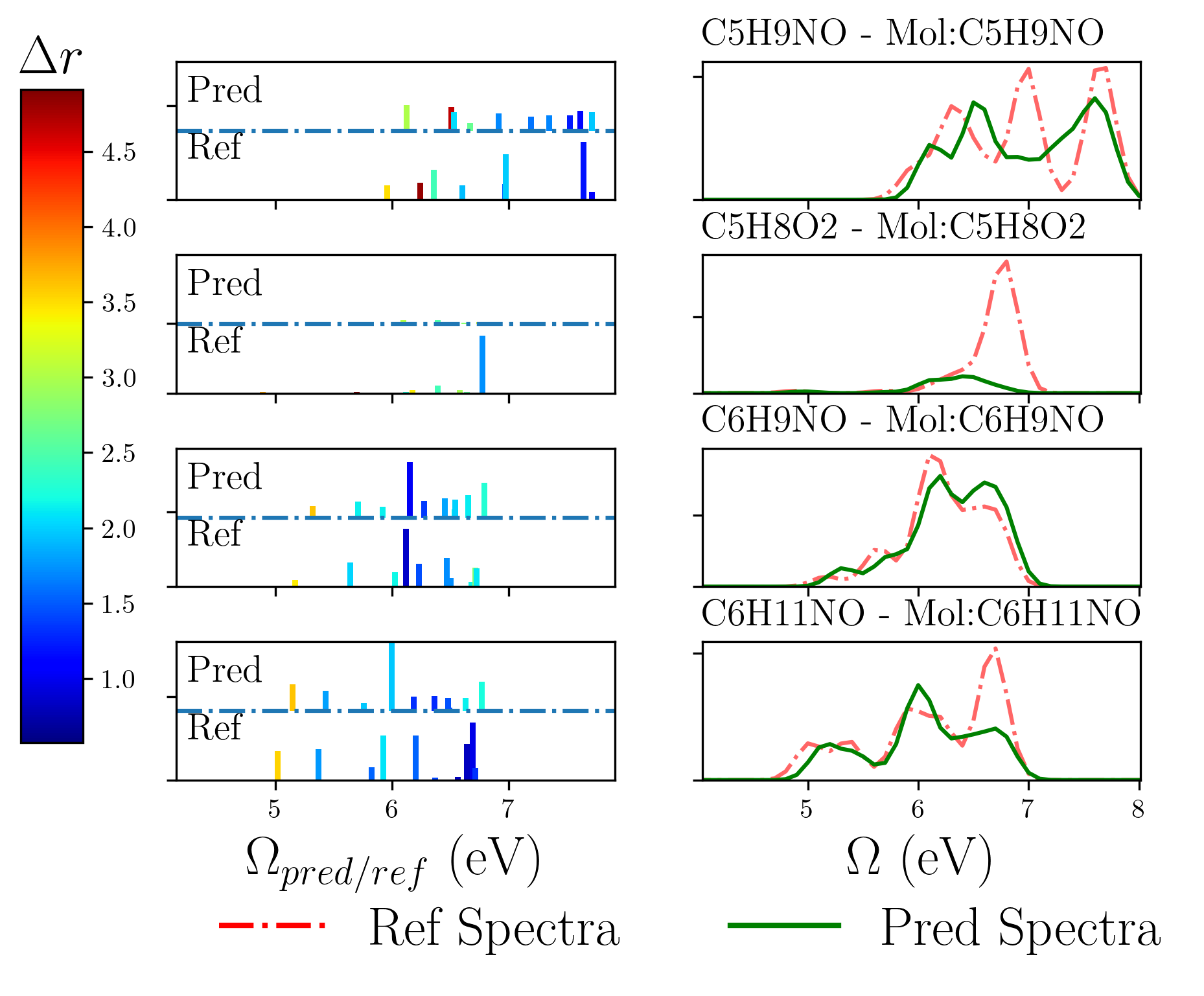}} &
{\includegraphics[scale=0.5]
 {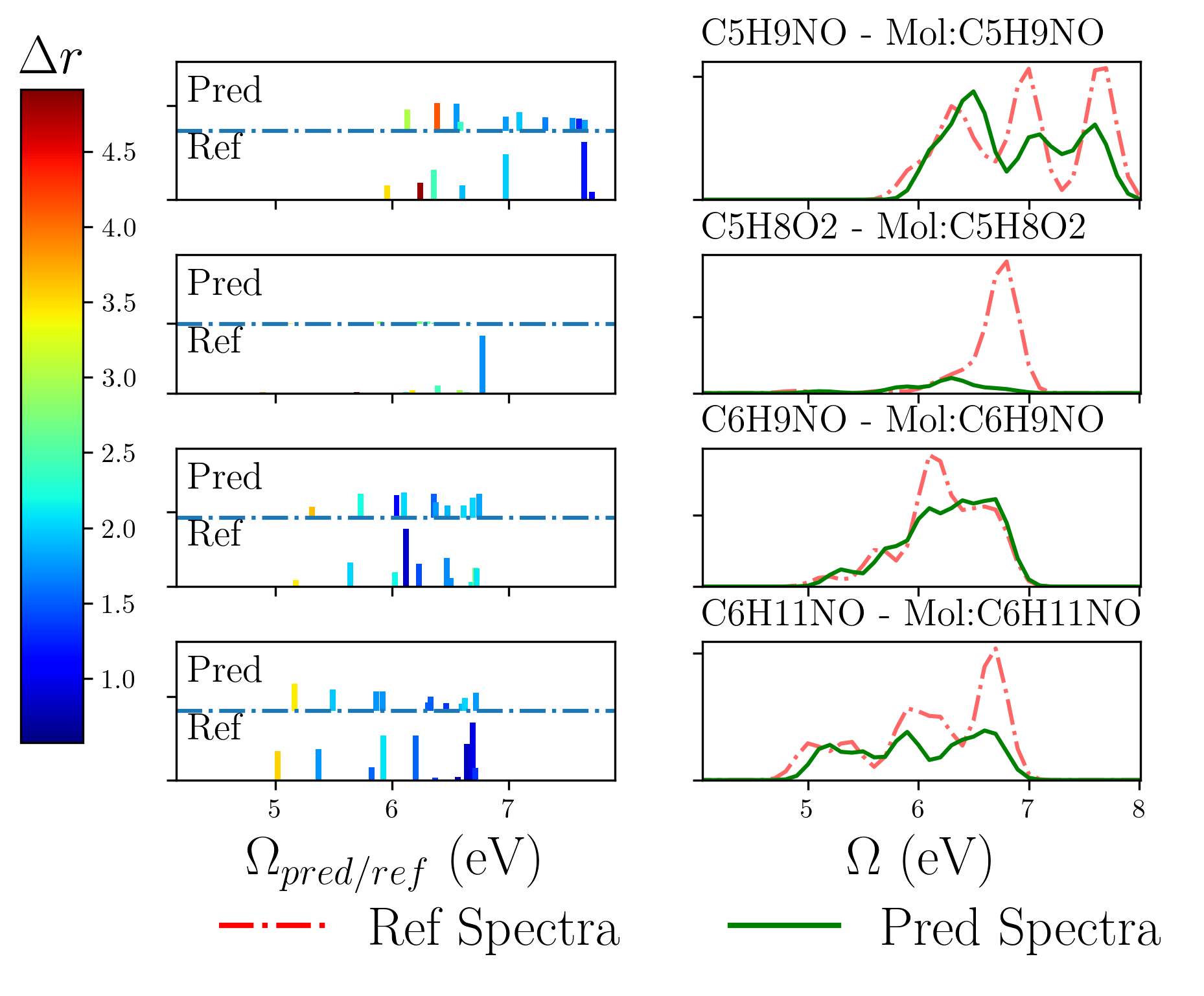}} \\
\end{tabular}
\caption{Discrete and broadened excitation spectra obtained with CNN-2D (a) and MLP-2D (b) for some example molecules contained in the control group. On the \emph{X} axis are the calculated excited state energies $\Omega$ and on the \emph{Y} axis appear their oscillator strengths $f_I$. Green curves represent reconstructed spectra from NN predictions, while the red ones represent those from PBE0-CASIDA calculations.
}\label{fig:brdn_spectra_sel_mol}
\end{figure}

In order to easily visualize the agreement of the models for predicting the optical response of a molecule we must also look at the absorption spectra. Figure \ref{fig:brdn_spectra_sel_mol} shows some examples of reconstruction for discrete and broadened absorption spectra (More examples can be found in the Supplementary Information). 

The discrete spectra is represented as impulses positioned at the specific excitation energy of a given state. Their heights are proportional to the calculated oscillator strength. The shown color of each impulse represents the index for charge transfer character according the color scales of $\Delta \boldsymbol{r}$ at left. It is well known that LDA functionals tend to underestimate excitation energies between Kohn-Sham's orbitals when they involve a charge transfer process. This is avoided in TDDFT calculations because the consideration of a fraction of the exact Hartree exchange potential for hybrid functionals, such as PBE0. If we look at the spectra of C5H9NO, the first excited state energy and nature in Figure \ref{fig:corr_examp}, that was obtained with simpler LDA calculation results used for NN learning can be compared with the predicted results of Figure \ref{fig:brdn_spectra_sel_mol}. A switch on the charge-transfer character of the first excited state appears together with a blue shift of the energy in the prediction. It means that even being our models fed with rough LDA calculated properties, the learning process appears to add the effect of the exchange-correlation, being this one of the more time demanding part when computing spectroscopic properties by TDDFT routines. It could be very significant to save computational resources and time for excited state predictions of large molecules.

Besides, the broadened spectra shown in Figure \ref{fig:brdn_spectra_sel_mol} have been reconstructed as a sum of Gaussians centered at the excitation energy which area is proportional to the oscillator strength. The broadening factor has been chosen to be $0.15$ eV  at the  half  width at the half maximum (HWHM). We used the cross-correlation between the normalized spectra, the curve shift and the curve area difference for comparison metrics between reference and predicted broadened spectra, because, the relevant property in spectroscopy is usually the relative intensity instead of its absolute value. Our models sometimes have difficulties to distinguish between closely laying excitations and can produce a switch between states. However, the broadening procedure enables us to mitigate this error by producing a good estimation of the continuum absorption spectra.

\begin{figure}[!ht]
\begin{tabular} {c c}
(a) CNN-2D &  (b) MLP-2D  \\ 
{\includegraphics[scale=0.55]
 {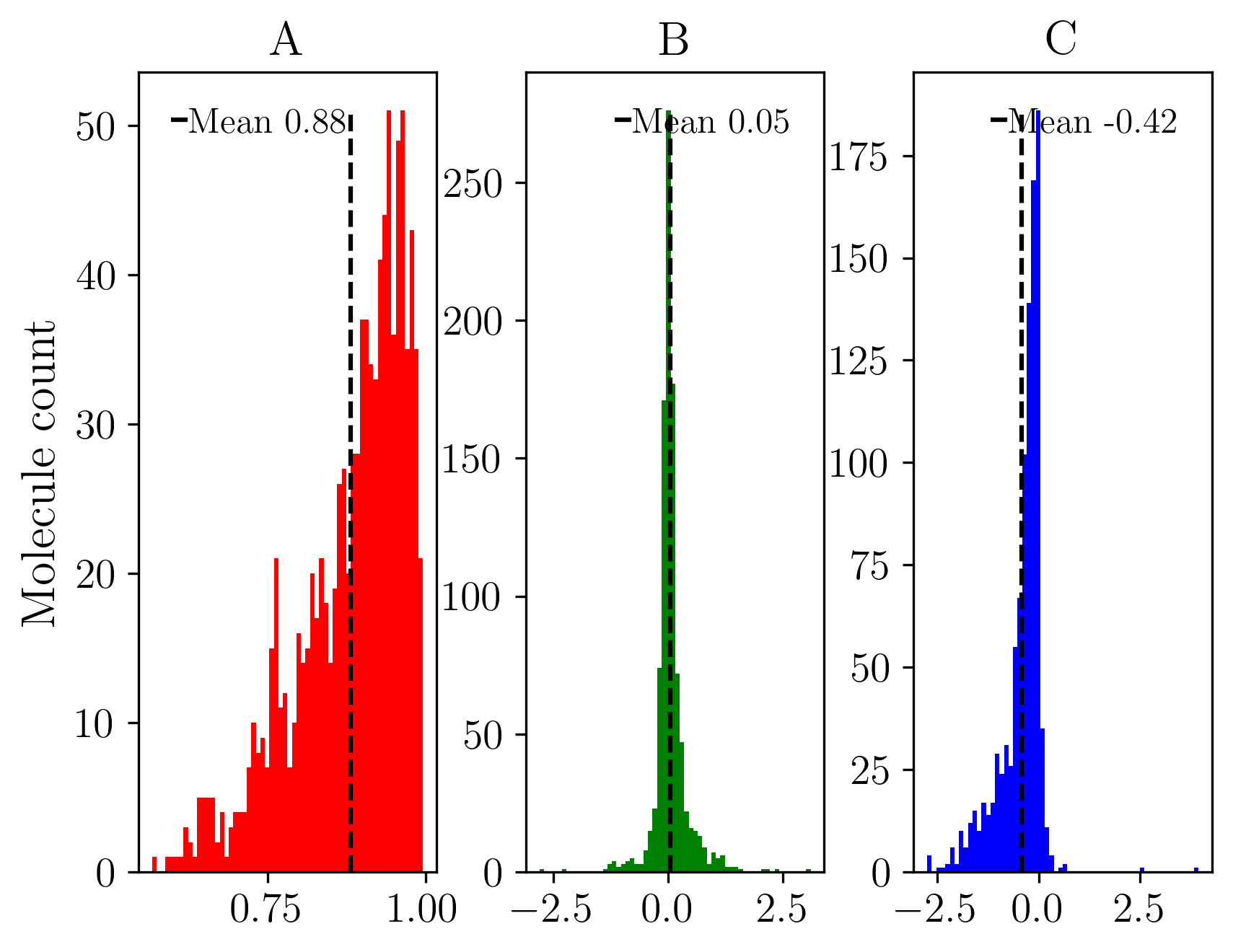}} & 
{\includegraphics[scale=0.55]
{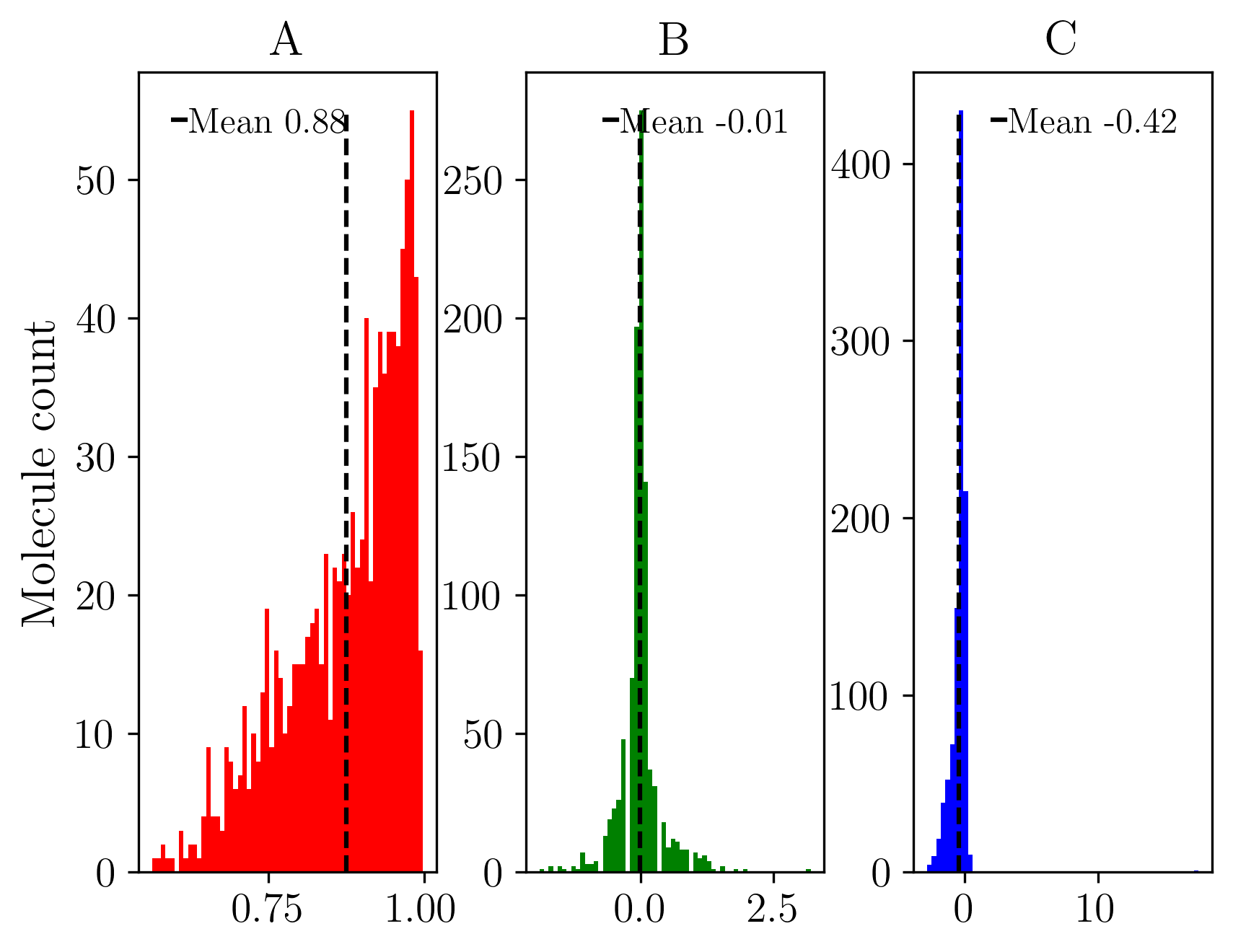}}
\end{tabular}
\caption{Distribution of the different metrics used to evaluate the prediction of the broadened spectra: (A, red) cross-correlation, (B, green) area under the curve and (C, blue) curve shift, for (a) CNN-2D (b) MLP-2D.}
  \label{fig:corr_hist}
\end{figure}

Figure \ref{fig:corr_hist} shows the distribution of the parameters used to evaluate the obtained broadened spectra. Both models shows a very good distribution of the cross-correlation values having most than 93\% situated above a value of 0.90. In addition, almost all tests produces an area under the curve close to the 0 values, which indicates that the number of electrons is conserved in the prediction. Regarding the shift of the predicted broadened spectra, we observe that in spite of the fact the great majority presents just a slight shift, both models tend to produce a small red-shift of the absorption spectra when comparing with the PBE0-CASIDA results.

\begin{figure}[!ht]
\begin{tabular}[c]{c c c}
{} & CNN-2D (a) & {} \\
{\scalebox{0.3}{%
    \includegraphics[]
    {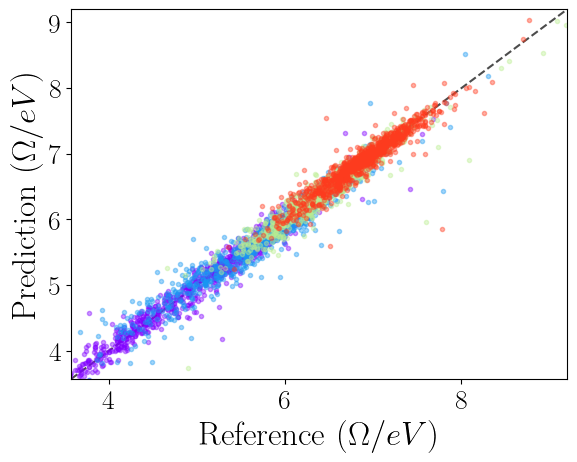}}} &
{\scalebox{0.3}{%
    \includegraphics[]
    {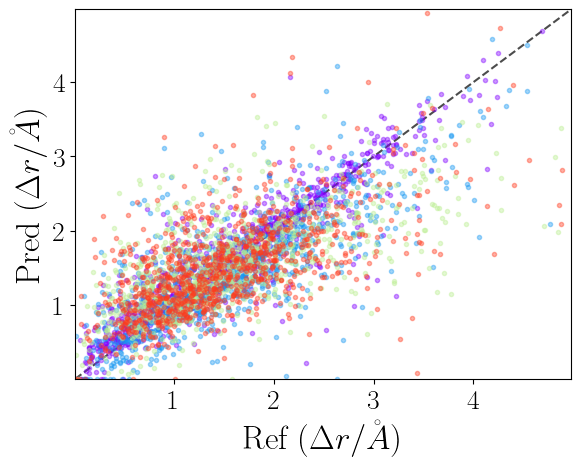}}} &
{\scalebox{0.3}{%
    \includegraphics[]
    {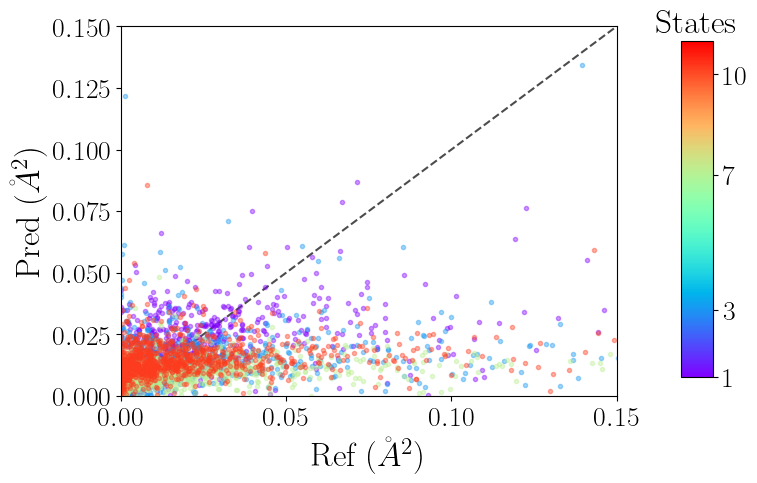}}} \\
{} & MLP-2D (b) & {} \\
{\scalebox{0.3}{%
    \includegraphics[]
    {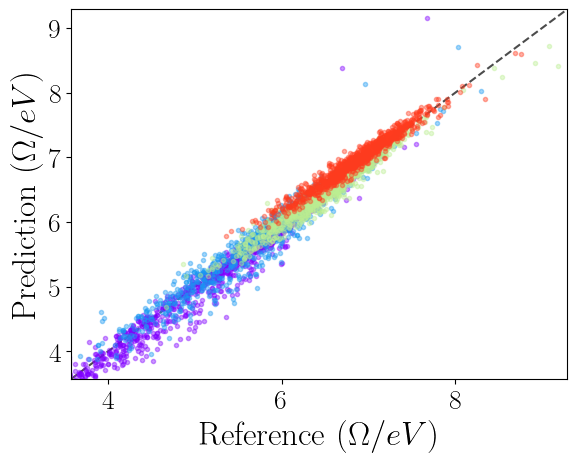}}} &
{\scalebox{0.3}{%
    \includegraphics[]
    {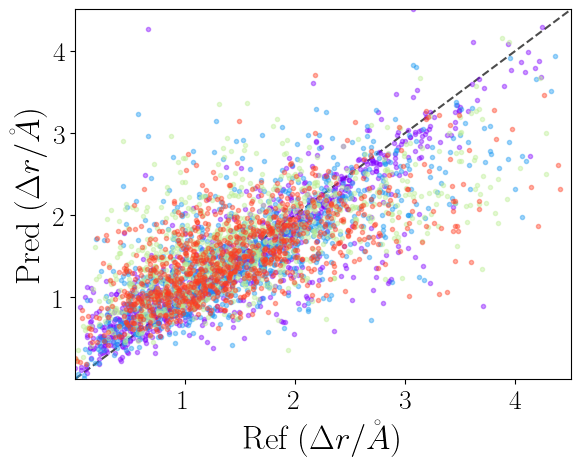}}} &
{\scalebox{0.3}{%
    \includegraphics[]
    {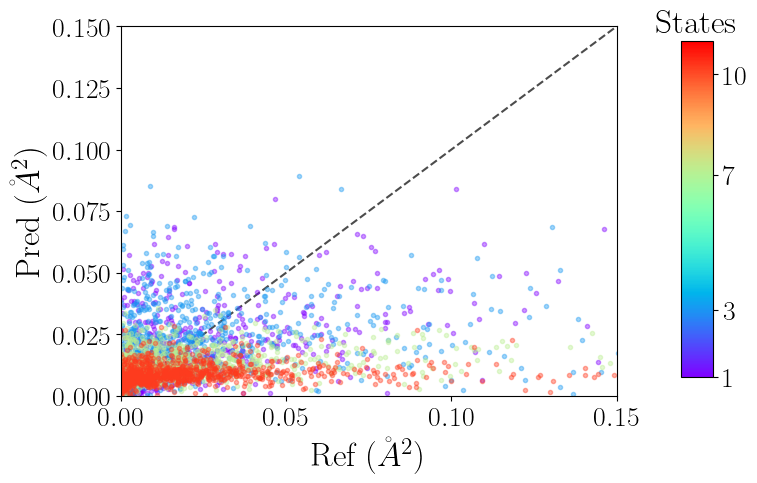}}} \\

\end{tabular}
    \caption{Correlation graphics between predicted and calculated values for best models (CNN-2D (a) and  MLP-2D (b). From left to right are represented: excitation energies ($\Omega$), charge-transfer coefficients ($\Delta r$) and transition moments ($\mu^2$).  The color of points correspond to the state number as ordered from the lowest energy according the scale at right.}
  \label{fig:corr_err}
\end{figure}

Figure \ref{fig:corr_err} shows the correlation graphics between the predicted values and the reference for each of the spectroscopic properties analysed in this work. We can see that our models produce a very good correlation of the excitation energies for all excited states. A better correlation is observed for higher excitation energies, which can be attributed to a higher density of states found in the database at such frequencies. Regarding the charge-transfer index, we see that the CNN-2D model produces better correlation for the low-laying states while it loose this correlation for higher states. It seems that both models hardly reproduce the proper transition dipole moments, being the CNN-2D model slightly better for the low-lying states. A possible source of error can be attributed to the diverse distribution of the values that increases the complexity of the learning process and/or to  intrinsic inconsistencies of the theoretical calculations of transition dipole moments when Kohn-Sham's virtual orbitals are involved. Other tests performed by increasing the size of our learning set including up to $15k$ molecules, suggests that enlarging the database can improve the correlation of the  $\Delta{\boldsymbol{r}}$ for higher excitations, but just produces an slight improvement on the transition dipole moments.

\section{Conclusions}

The accurate knowledge of the spectroscopic properties of molecules has been of great interest since long ago for academical as well as industrial sectors. The high computational cost of the quantum chemistry/physics techniques, mostly for large molecules, and the lack of an extended experimental database, as well as the increase on the reliability of the machine learning and artificial intelligence methods, are inviting researchers to apply those techniques for predicting these physical properties. Nevertheless, the major difficulty usually relies on finding the proper descriptors being able to correlate with properties of interest.

As mentioned above, the main objective of this work is to find adequate molecular-electronic descriptors to be used with proper NN models to predict the theoretical absorption spectra for a group of small organic molecules. We prove that the combination of certain selected electronic properties ($\Delta \epsilon_{ia}$, $R_{ia}$, $\mu_{ia}^2$) resulting from low cost \emph{ground state LDA} calculations appear as good descriptors for the prediction of such spectroscopic properties at a higher level of theory (e.g. PBE0 functional). Besides, we demonstrate that a simple optimised Convolutional Neural Network (CNN-2D) as well as a Multi Layer Perceptron (MLP-2D) network can learn to supply the exchange correlation correction required for predictions going from LDA to PBE0 levels of theory. 

Previous works were focused on the prediction of the first excited state\cite{ml_6}, or on the density of states near to the LUMO \cite{ml_35} by using only geometrical or spacial descriptors. In this work, we demonstrate the need of an electronic descriptor not only to extend the prediction of the excitation energies at the chemical accuracy, but also to give information about their charge-transfer character. Oscillator strength values proved to be the most challenging property for our models. Although we demonstrate an enhancement on the prediction of the low-laying excitation probabilities when the training set is augmented, the transition dipole moments for high energy excitation remained poorly correlated. Different sources of error that can be addressed for this problem are discussed.  

The natural next step is to provide even more fundamental properties as descriptors to the Neural Network. However, in progress work reveals that the use of the same properties coming from unoptimized Linear Combination of Atomic Orbitals (LCAO) calculations, typically used to begin a ground state calculation, requires a more complex network optimization to overcome the big gap between a LCAO level of theory and that of PBE0 hybrid functional calculations.

All data from the calculations done in this work have been stored in the database of the Novel Material Discovery (NOMAD) project.\cite{ml_146} , and can be downloaded: LDA from \url{https://dx.doi.org/10.17172/NOMAD/2021.10.18-2} and PBE0 from \url{https://dx.doi.org/10.17172/NOMAD/2021.10.18-3}.

\begin{acknowledgement}

This work was supported by the European Research Council (ERC-2015-AdG694097), the Cluster of Excellence 'CUI: Advanced Imaging of Matter' of the Deutsche Forschungsgemeinschaft (DFG) - EXC 2056 - project ID 390715994, Grupos Consolidados (IT1249-19) and  the SFB925 "Light induced dynamics and control of correlated quantum systems”. We kindly recognize the partial support of the project ID PN223LH010-002 "Inteligencia Artificial Aplicada. Espectroscopía y Bioactividad" of the Cuban Ministry of Science, Technology and Environment as well as the overall support given to LAMC by the Universidad de La Habana and the Donostia International Physics Center. 

\end{acknowledgement}

\begin{suppinfo}

Additional data and NN descriptions.

\end{suppinfo}

\bibliography{ml_spectra.bib}
\end{document}



\subsection{Description of the Bayesian Optimization}
\begin{figure}
\centering
\subfloat[CNN-2D]
        {\includegraphics[scale=0.5]
        {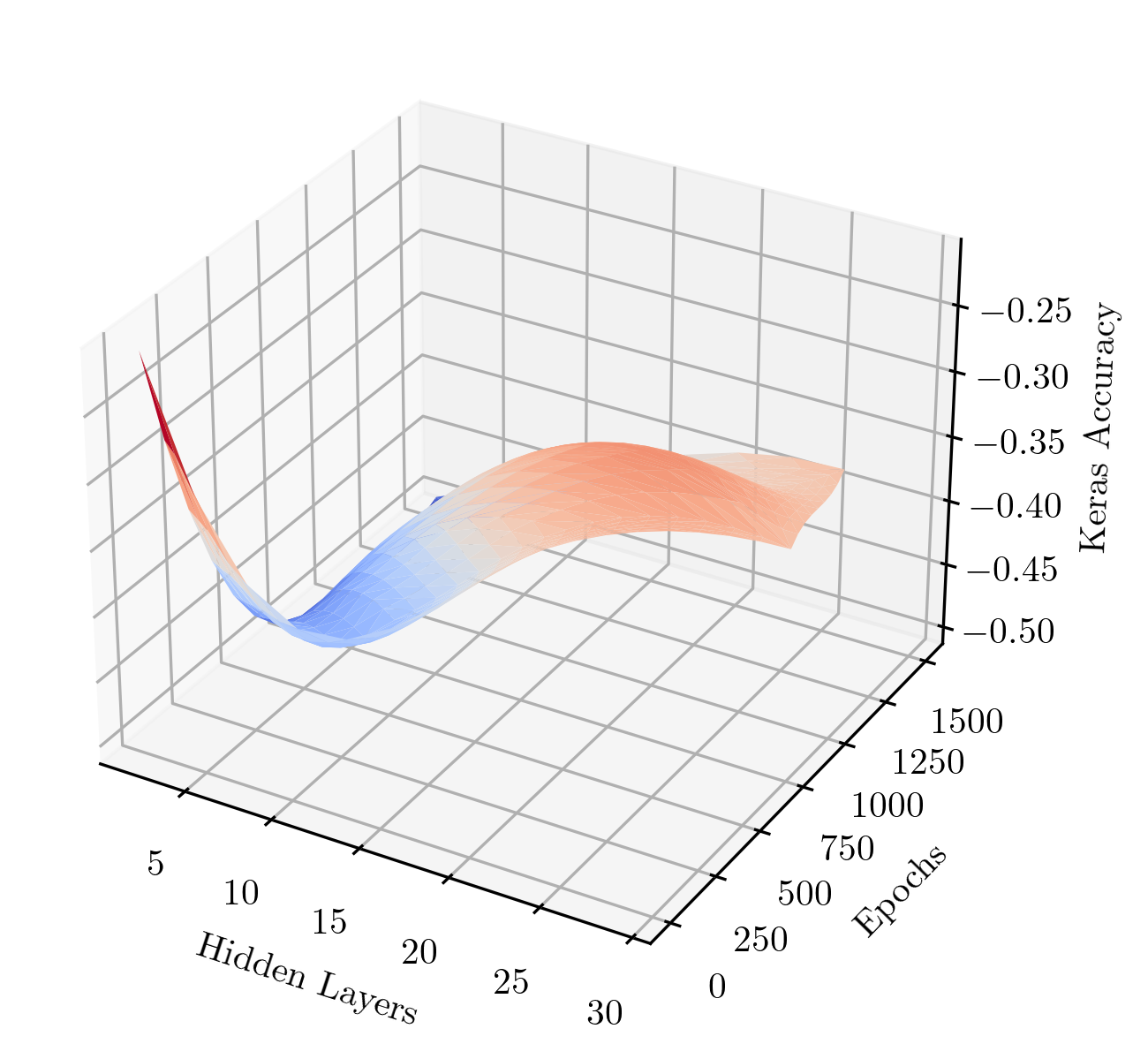}}
\subfloat[MLP-1D]
        {\includegraphics[scale=0.5]
        {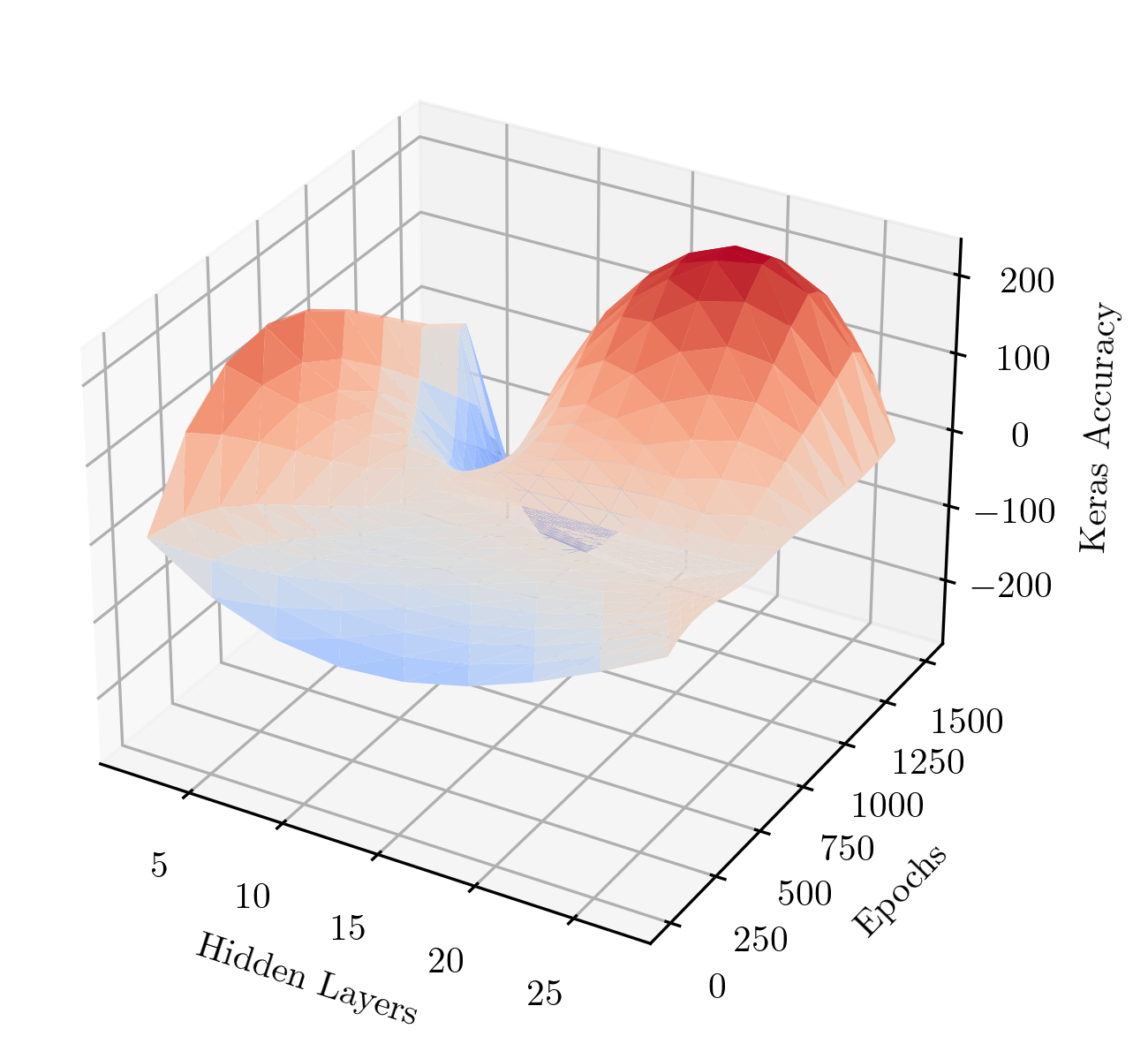}}\\
\subfloat[MLP-2D]
        {\includegraphics[scale=0.5]
        {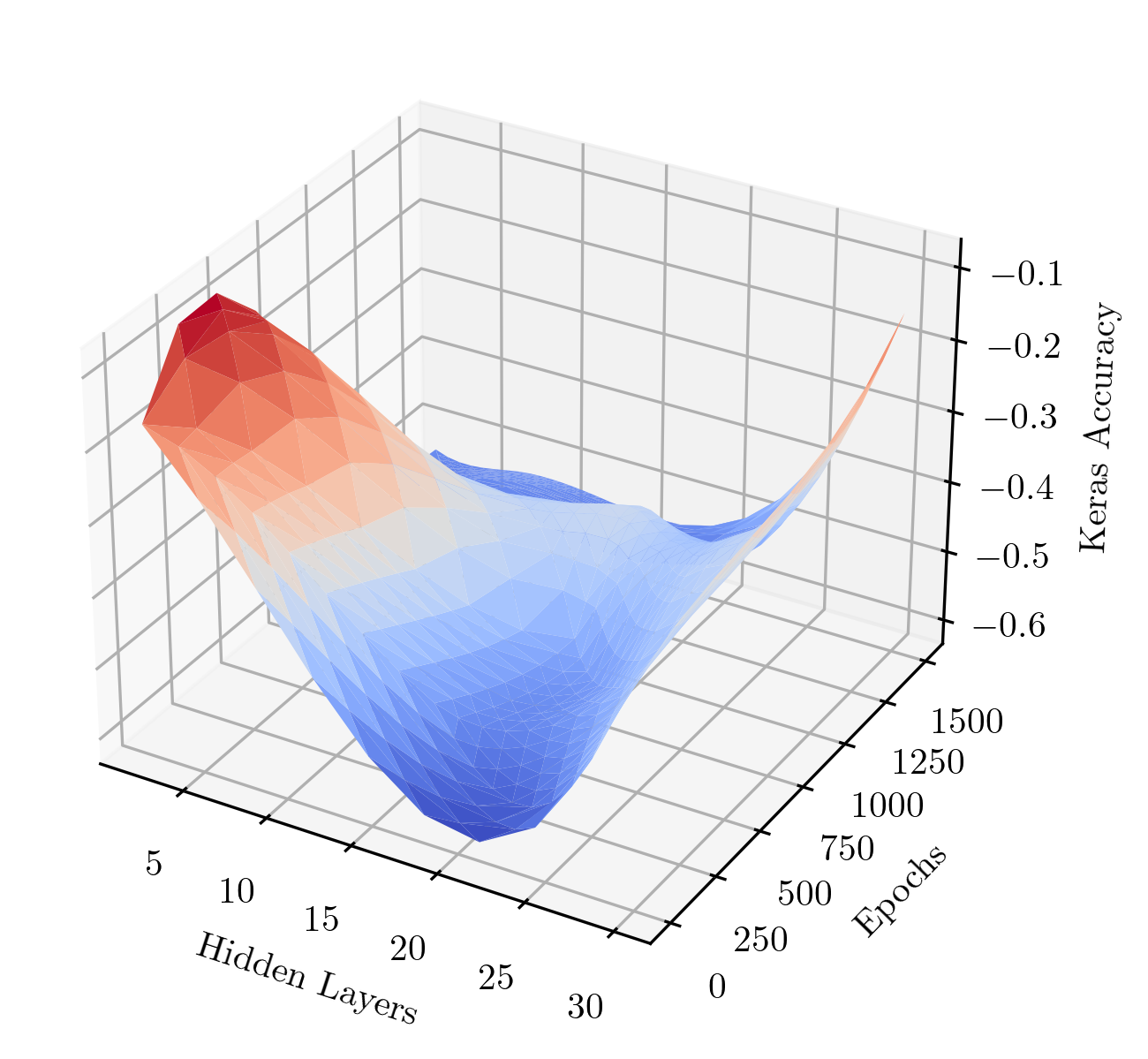}}
\caption{Accuracy vs (Hidden-Layers, Epochs) of our models for the Bayesian optimization. Using LDA-GS as descriptor and PBE0-CASIDA as target property. In \emph{\color{red}{red}} those combination where the models gives the worst resultes, in \emph{\color{blue}{blue}} the best combination of hyper-parameters.}
\label{fig:bay_hypersurf}
\end{figure}

The number of hidden layers and epochs for each topology to be determine by Bayesian Optimization. For epochs we select the values $1, 100, 500,$ $1000, 1500$ and for hidden layer the values $1, 2, 5, 10, 20, 30$. The combinations of this values form the search space for the Bayesian Optimization. Figure \ref{fig:bay_hypersurf} shows the hyper-surface resulting of the evaluation of the models. The red spots are where the models gives the worst possible results and the blue areas are where the combination of
parameter gives the best possibles results. Those blue areas point to our selected combinations of hidden-layers and epochs. The metric used was the \emph{Accuracy} implemented by Keras.

Table \ref{tab:hy_param_list} shows the rest of hyper-parameter values selected to construct out models. The model where constructed using Keras with TensorFlow, the names coincides with those implemented in by the frameworks named before.

\begin{table}
 \caption{List of hyper-parameter for our models and how their values were
 determine.}
\resizebox{15cm}{!}{
\begin{tabular}{cccccccc}
\hline
Model & Activation & Number of Neurons per Hidden-layer & Optimizer & Loss Function & Kernel size \\
\hline
MLP-1D & eLU/ReLU(negative slope = 0.01)  & 60 & Adams& MSE & - \\
MLP-2D &  eLU/ReLU(negative slope = 0.01)  & 25 & Adams  & MSE & - \\
CNN-2D &  eLU/ReLU & 60 (Filter)  & Adams  & MSE & 20
\end{tabular}
}\label{tab:hy_param_list}
\end{table}

\subsection{Reconstruction of the discrete absorption spectra.}

\begin{figure}
\centering
\subfloat[CNN-2D]
        {\includegraphics[scale=0.5]
        {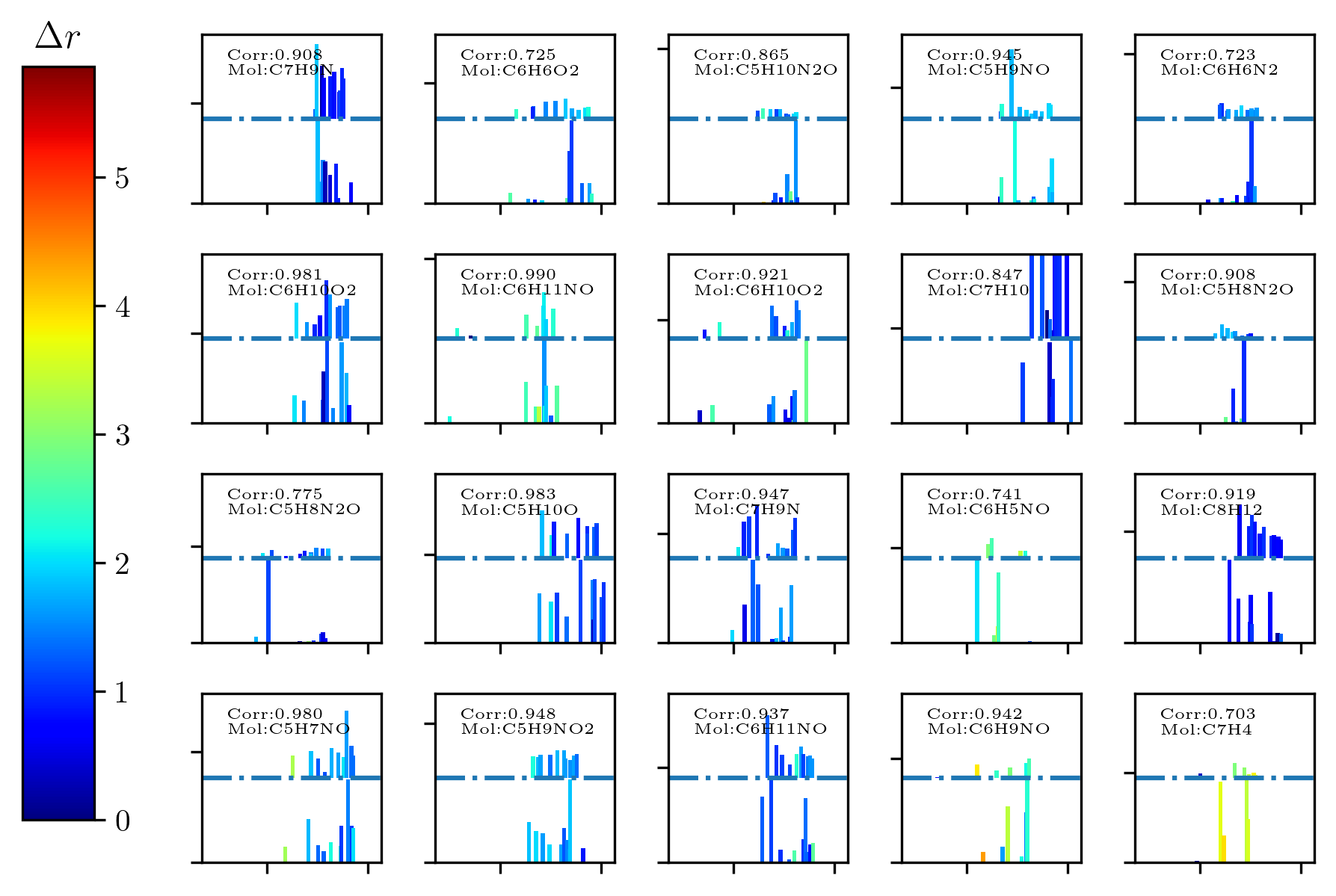}}
\subfloat[MLP-2D]
        {\includegraphics[scale=0.5]
        {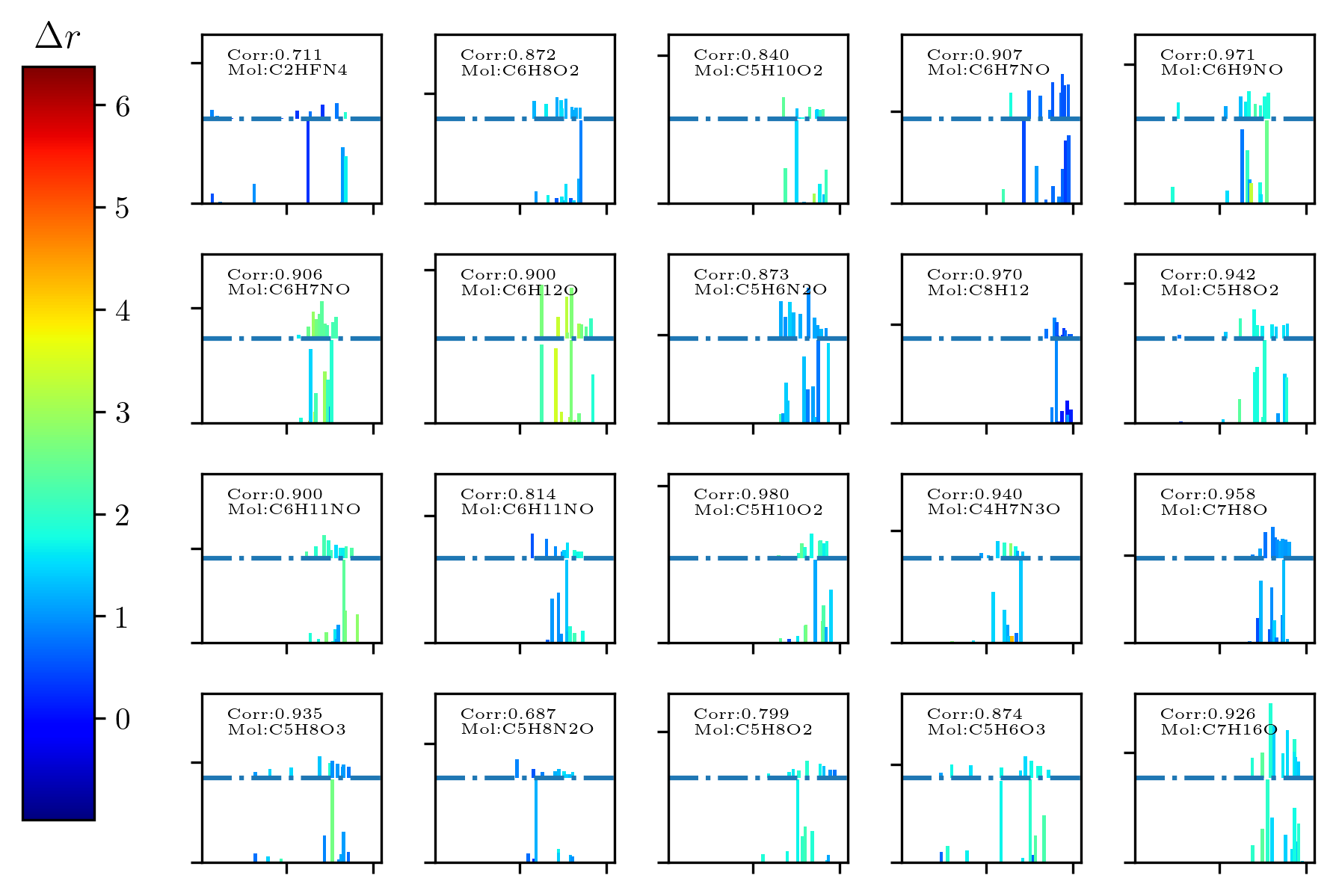}}
\caption{Discrete absorption spectra reconstruction. The color scale shows the intensity  of the transition measured by the $\Delta r$ metric. Each image shows a molecule, were the upper reconstruction is the prediction a the lower reconstruction is the reference. The descriptor used was extracted from LDA-GS and the target properties from PBE0-CASIDA.}
\label{fig:bash_recons}
\end{figure}

As an example of multiple molecules discrete absorption spectra, Figure \ref{fig:bash_recons} shows 20 molecules for CNN-2D and MLP-2D models. In both cases the function $log$ was used as a pre-possessing method.

\begin{figure}[!ht]
\begin{tabular} {c c}
(a) CNN-2D &  (b) MLP-2D  \\ 
{\includegraphics[scale=0.5]
 {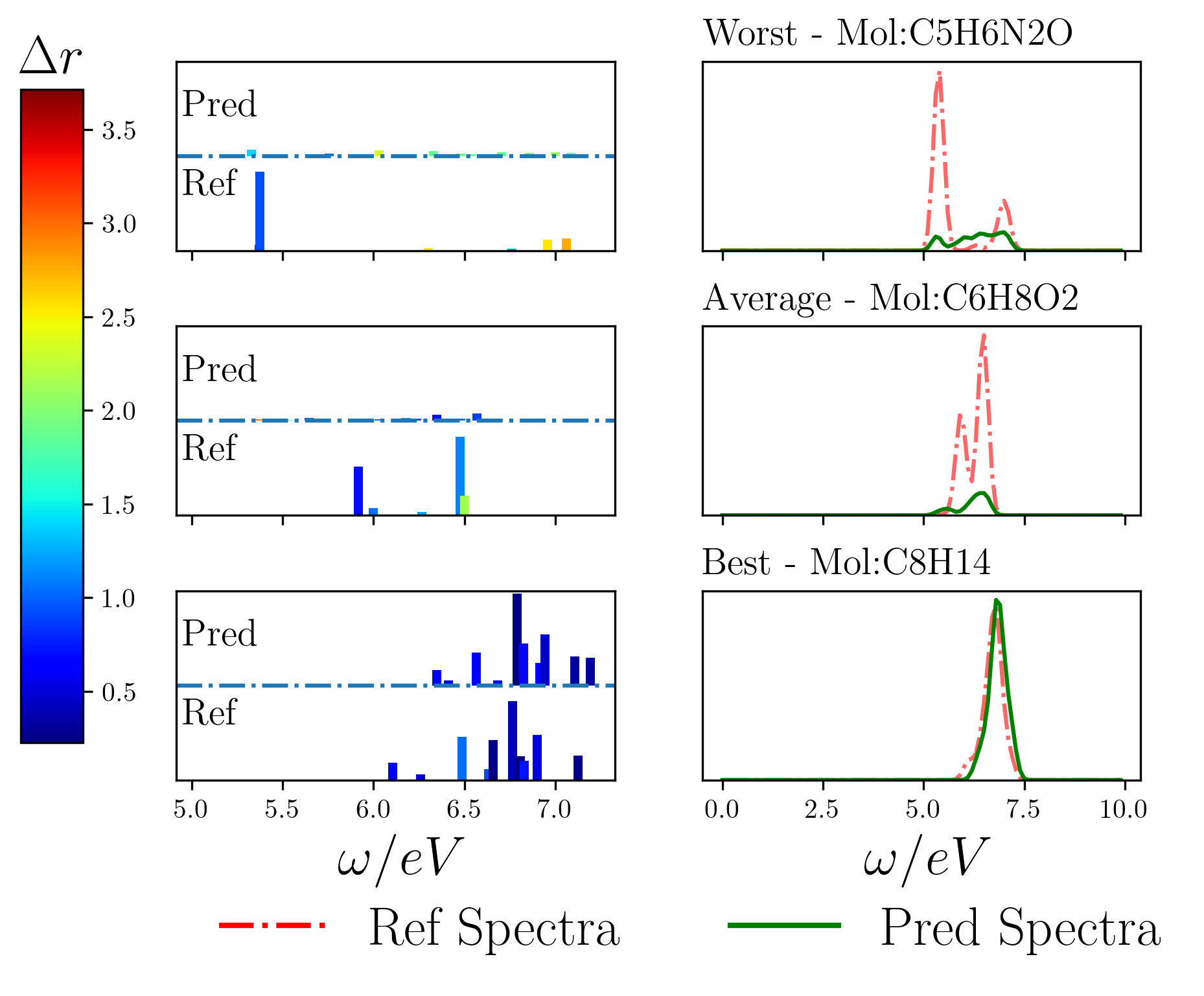}} & 
{\includegraphics[scale=0.5]
{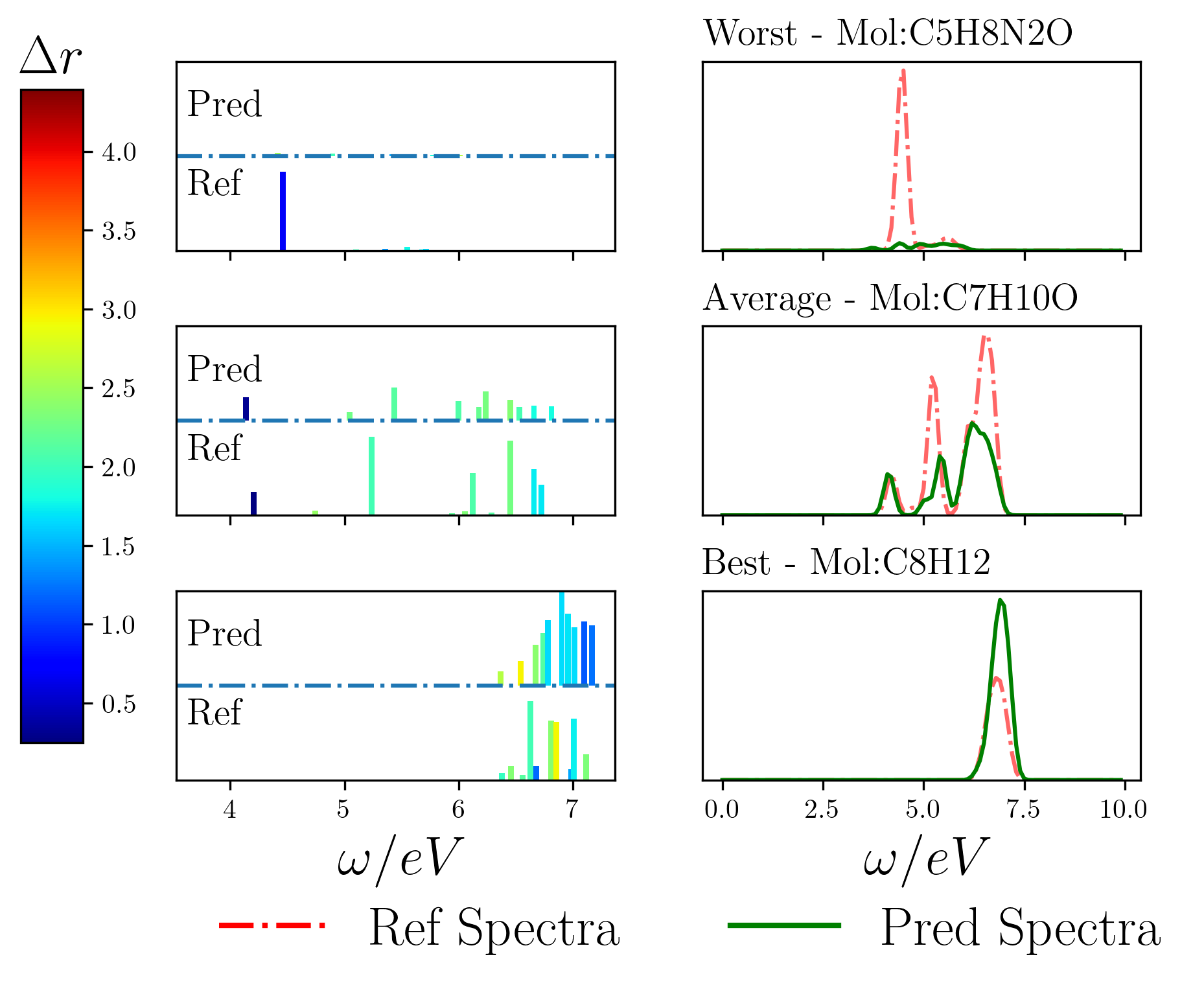}}
\end{tabular}
\caption{Discrete and broadened excitation spectra for the best, mean and worst examples: (a) CNN-2D (b) MLP-2D. On the \emph{X} axis are the excites states $\omega$ and on the \emph{Y} axis the oscillator strengths $f_I$. Green curves represent reconstructed spectra from predictions, while the red ones represent the reference reconstructed spectra from PBE0-CASIDA calculations.
}\label{fig:brdn_spectra}
\end{figure}

For simplicity Figure \ref{fig:brdn_spectra} show the best, mean and worst spectra reconstruction. 

